\documentclass[preprint,aps,draft]{revtex4}
\usepackage{epsf}
\begin{document}
\title{Comment on "Universal effect of excitation dispersion on the 
heat capacity and gapped states in fluids"}

\author{Taras Bryk$^{1,2}$,
        No\"el Jakse$^{3}$,
        Ihor Mryglod$^{1}$,
        Giancarlo Ruocco$^{4,5}$,
        Jean-Fran{\c c}ois Wax$^{6}$}

\affiliation{ $^1$ Institute for Condensed Matter Physics,National
Academy of Sciences of Ukraine,\\UA-79011 Lviv, Ukraine}
\affiliation{$^2$Institute of Applied Mathematics and Fundamental
Sciences,\\Lviv National Polytechnic University, UA-79013 Lviv,
Ukraine } 
\affiliation{$^3$ Universit\'e Grenoble Alpes, CNRS, Grenoble INP, SIMaP, F-38000
        Grenoble, France}
\affiliation{$^4$
Center for Life Nano Science @Sapienza, Istituto Italiano di
  Tecnologia, 295 Viale Regina Elena, I-00161, Roma, Italy}
\affiliation{$^5$ Dipartimento di Fisica, Universita' di
Roma "La Sapienza", I-00185, Roma, Italy} 
\affiliation{$^6$ Laboratoire de Chimie et de Physique A2MC, Universit\'e de Lorraine,
        Metz, 1, boulevard Arago 57078 Metz Cedex 3, France}

\date{\today}
\begin{abstract}
We discuss the validity of recent results in [Phys. Rev. Lett. 125, 125501 (2020)] on 
an universal relation between the heat capacity and dispersions of collective excitations
in liquids. 
\end{abstract}

\maketitle

It is well known from the textbooks\cite{Han} that the heat capacity of liquids is defined by 
thermal and heat density fluctuations. In a recent Letter\cite{Kry20} the authors made a 
 claim that 
there exists a universal connection between the heat capacity of liquids and the value of the 
propagation gap for shear waves: "universal bilinear dependence of heat capacity on q-gap width". 
However, there is no theory in support of the claimed effect of overdamped short-wavelength 
transverse collective modes on specific heat $C_V$. These are in real liquids the short-time 
ones and 
cannot store the energy that is required for their contributions to $C_V$. On the contrary,
the standard fluctuation theory \cite{Sch66} states that the heat fluctuations are orthogonal 
to the fluctuations of transverse components of mass current and of stress tensor.
In Fig.1 the specific heat as a function of wave number according to \cite{Sch66,Cop75,
Bry13}
is shown within solely $q$-dependent heat density fluctuations. The $q$-dependence is
in very good agreement with the macroscopic $C_V$ obtained via standard heat fluctuations of the
whole system \cite{Sch66}. No contributions from non-hydrodynamic transverse excitations to specific
heat exist in this standard approach.
Moreover, it exists at least a handwaving counter example to the claimed "universal behavior" 
\cite{Kry20}: in hard-sphere fluids the heat capacity is $C_V=1.5 k_B$ for
any density, while the propagation gap in the dispersion of shear waves is strongly changing 
with density \cite{Bry17_hs}.

Another critical comment deserves the methodology of estimation of dispersion of longitudinal (L)
and transverse (T) collective excitations, where the authors claim the effect of "anticrossing"
of the L and T branches. First, the general symmetry rules for correlations in simple liquids 
require that 
any canonically averaged local (with the same $q$ value) cross-correlation between L and T dynamic 
variables $\langle A^L(-q,t)A^T(q,t')\rangle\equiv 0$ is exactly zero, that means the L and T 
branches at the same wave number cannot repell each other. Second, the general models for explanation
of current spectral functions {\it must} contain contributions from slow relaxation processes 
like structural relaxation, which are absent in the used "two-oscillator model" and which are
extremely important in the region of de~Gennes slowing down of density fluctuations, i.e.
namely where in \cite{Kry20} the large "anticrossing" between L and T excitations was reported. 
In Supplementing material one can see how the fit-free generalized hydrodynamic theory,
which perfectly reproduces a large number of exact sum rules, allows to describe the current-current 
time correlation functions by proper fit-free contributions from dynamic eigemodes without any 
forbidden by the symmetry rules "anticrossing" of L and T modes.  Yet, the fit ansatz in 
\cite{Kry20} violates the exact relations for longitudinal and total current spectral functions
$C^L(q,\omega=0)\equiv 0$ and $C^{tot}(q,\omega=0)=2mk_BT\frac{\rho}{q^2\eta(q)}$, where 
$\eta(q)$ and $\rho$ are $q-$ dependent shear viscosity and mass density, respectively.
\begin{figure}\label{Fig1}
\epsfxsize=.40\textwidth {\epsffile{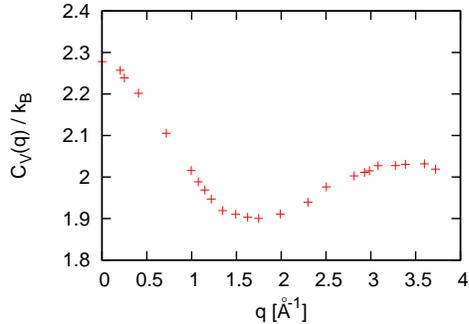}}
\caption{ Wave-number dependent specific heat $C_V(q)$ via solely heat fluctuations in
supercritical Ne at density $1600~kg/m^3$ and temperature $295~K$. 
}
\end{figure}

\section*{Supplementing material}

In order to check the results of \cite{Kry20} we performed molecular dynamics simulations (MD) 
using 4000 particles for Lennard-Jones fluids, potential parameters of which corresponded to Ne.
All the simulation setup was the same as in \cite{Bry17}. The shear waves (transverse propagating
modes) were observed only for the four highest densities of the studied supercritical Ne. The 
value of the gap $q_g$ for shear waves is shown in Fig.2a
by "plus" symbols, while "cross" symbols 
show the value of the Debye wave numbers, which were obtained as the $q_D=q_{max}/2$ with 
$q_{max}$ being the location of the main peak of the static structure factor $S(q)$.
In Fig.2b the specific heat from MD simulations is compared with a prediction from a
"phonon theory" based on non-damped excitations \cite{Fom18} (which was mentioned as a 
motivation in \cite{Kry20}) as a function of the ratio 
$q_g/q_D$.  It is seen from Fig.2b that the approach of non-damped ”phonons” in liquids
strongly overestimates the effect of short-wavelength collective modes on specific heat $C_V$.
\begin{figure}\label{Fig1S}
\epsfxsize=.45\textwidth {\epsffile{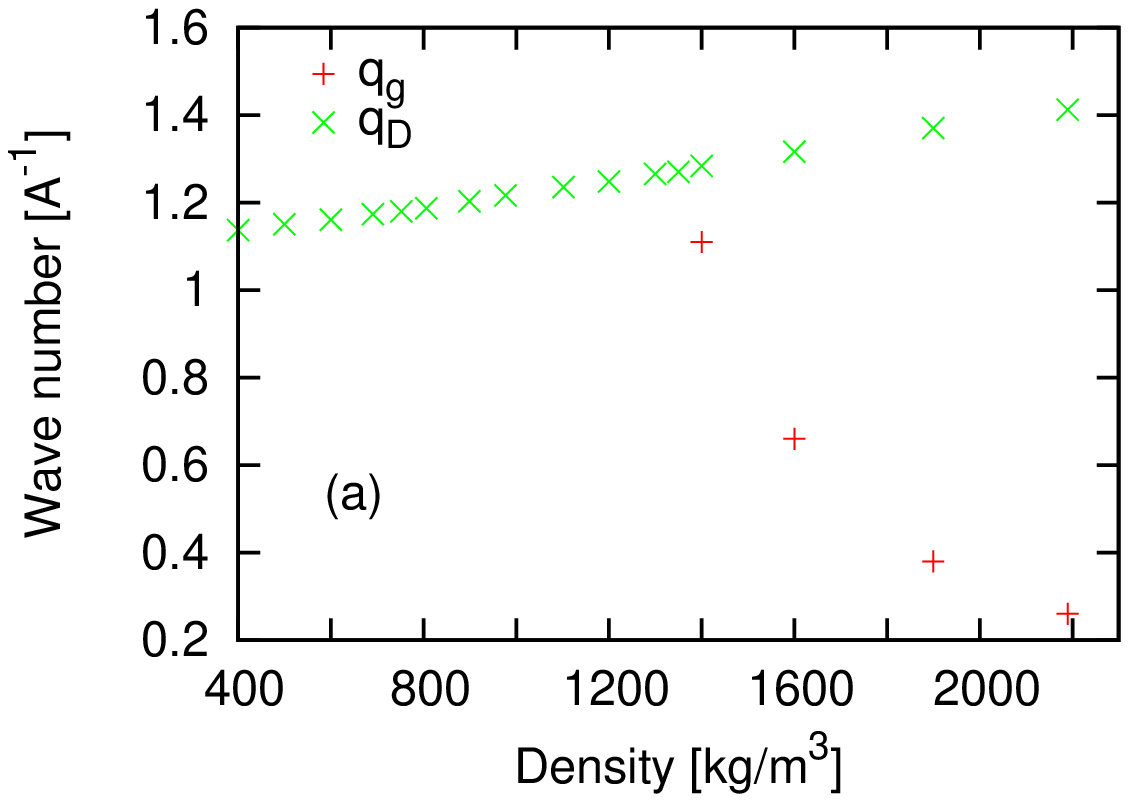}}
\epsfxsize=.45\textwidth {\epsffile{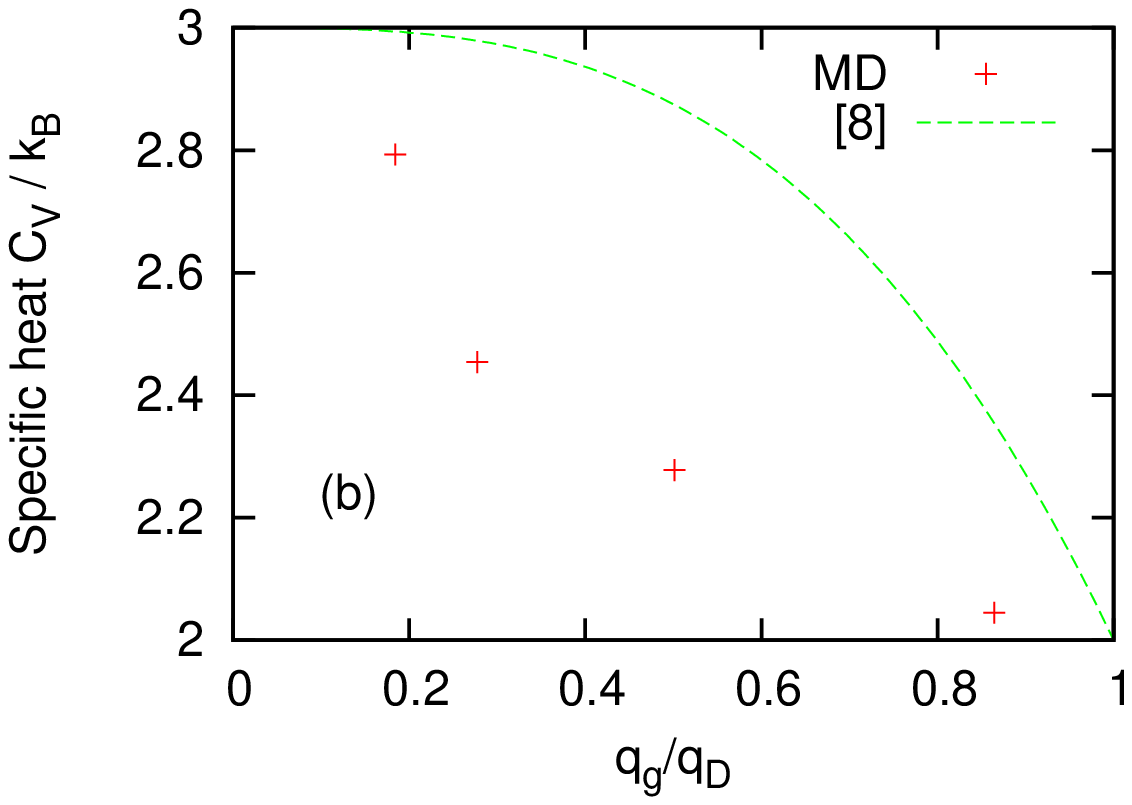}}
\caption{Density dependence of the $q$-gap and the Debye wave 
number (boundary of the first pseudo-Brillouin zone) for supercritical 
Ne at 295~K (a). 
Heat capacity vs $q_g/q_D$ as predicted by the "phonon theory of liquid thermodynamics" 
\cite{Fom18} (line) and directly calculated
via temperature fluctuations in MD simulations values in supercritical Ne at 295~K (b). 
}
\end{figure}

We performed an analysis of dynamic eigenmodes and their contributions to time correlation 
functions based on the eigenvalue problem for generalized hydrodynamic matrix. This approach
provides identical results both from kinetic theory with generalized Enskog operator \cite{deS88}
and from generalized hydrodynamics based on generalized Langevin equation \cite{Mry95}. The 
theoretical density-density (Fig.3a) and longitudinal mass current-current 
(Fig.3b) time 
correlation 
functions reproduce very nicely MD-derived functions without any fit within the 
five-variable thermo-viscoelastic
model\cite{Bry10}:
\begin{equation} \label{a5}
{\bf A}^{(5)}(q,t) = \left\{q(k,t), J^L(q,t), h(q,t),
\dot{J}^L(q,t), \dot{h}(q,t)\right\},
\end{equation}
where first three dynamic variables in the set (\ref{a5}) correspond to the fluctuations of 
the conserved quantities, while the two latter in (\ref{a5}) are the first time derivatives of 
the hydrodynamic variables, are orthogonal to them and 
describe fluctuations of the longitudinal component of stress tensor and of the heat current.
The thermo-viscoelastic set (\ref{a5}) of dynamic variables was applied to derivation of the 
$5\times 5$ generalized hydrodynamic matrix and finding its $q$-dependent eigenvalues $z_i(q)$ 
and corresponding eigenvectors.
 Contributions from the same eigenvalues $z_i(q)$ according to 
$$
F_{\alpha\beta}(q,t)=\sum_{i=1}^{5}G^i_{\alpha\beta}(q)e^{-z_i(q)t},\qquad \alpha,\beta=n,J^L
$$
with corresponding amplitudes $G^i_{\alpha\beta}(q)$ estimated via eigenvectors associated with 
the appropriate eigenvalues are shown in Figs.3c,d and give evidence what kind of eigenmodes 
must be taken into account in any approximate fit approach for dynamic correlations in liquids.
In Fig.3d one can see what should be the contribution from structural relaxation to the longitudinal
 mass current-current time correlation function, which however is absent in the proposed fit 
in \cite{Kry20}.
\begin{figure}\label{Fig2S}
\epsfxsize=.45\textwidth {\epsffile{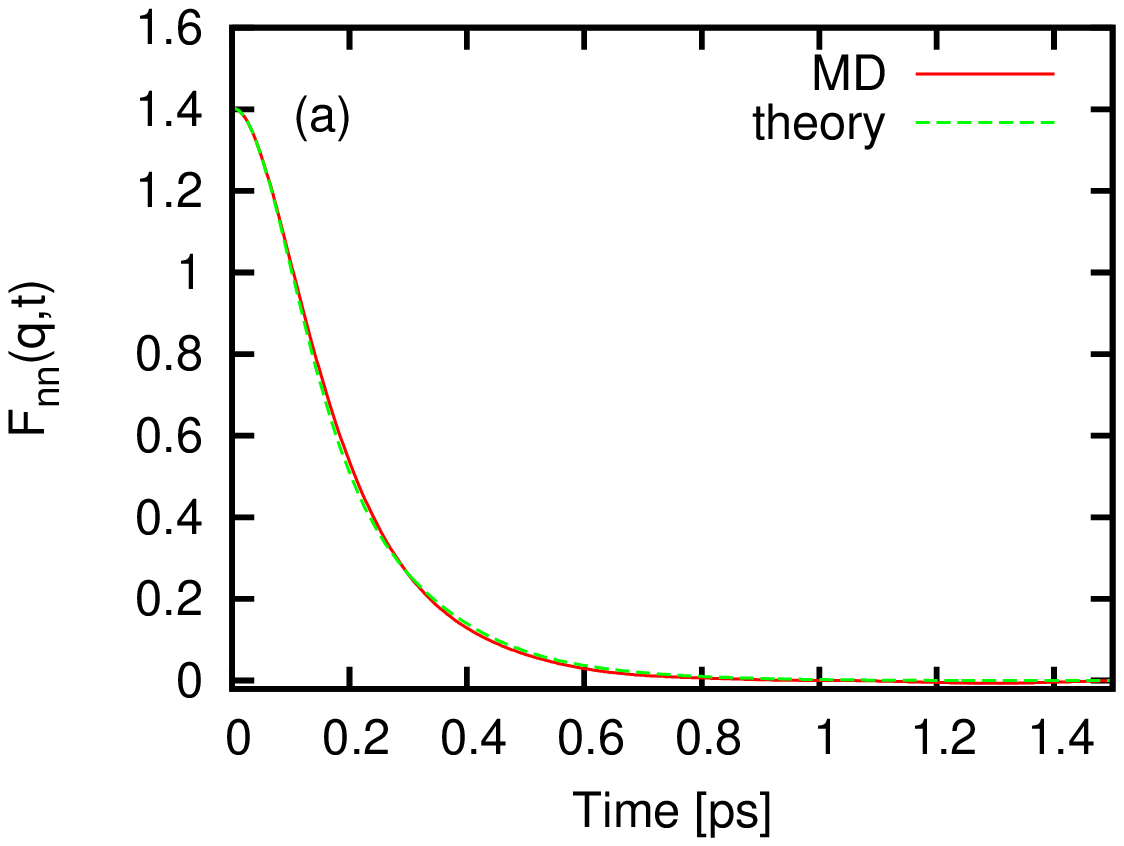}}
\epsfxsize=.45\textwidth {\epsffile{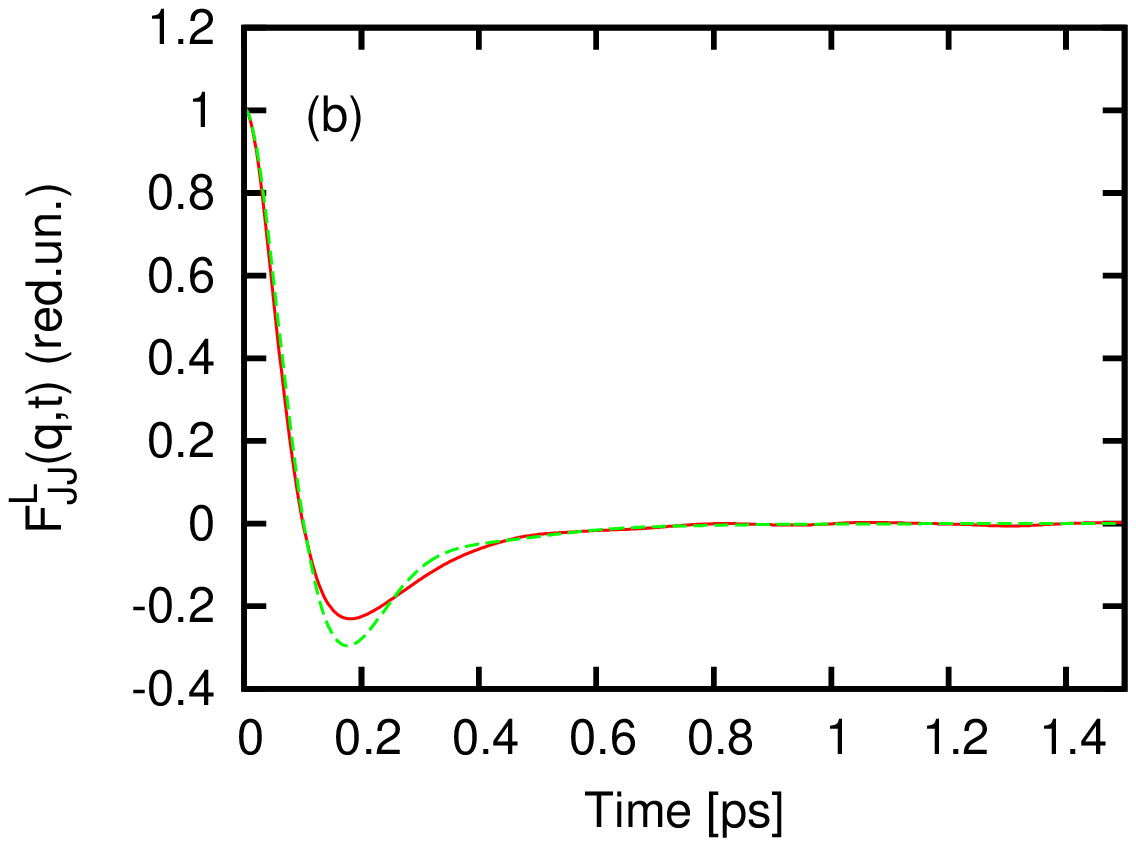}}

\epsfxsize=.45\textwidth {\epsffile{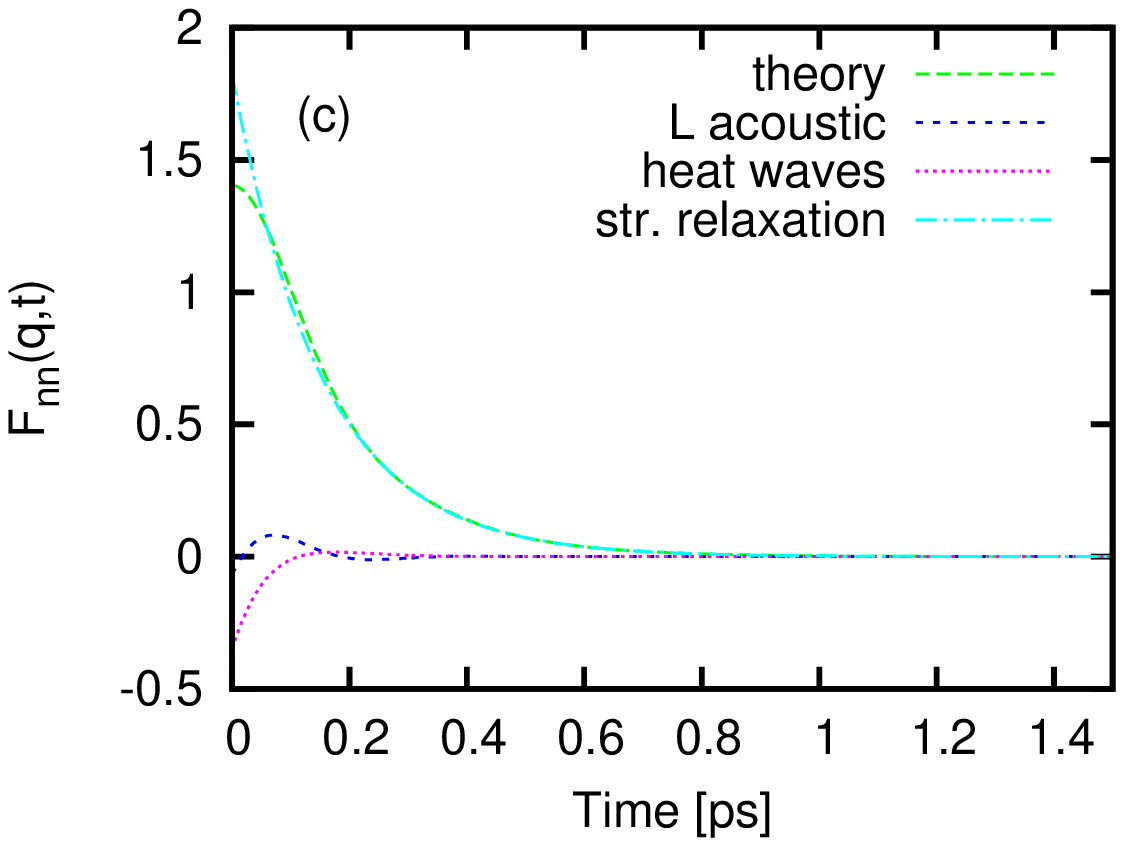}}
\epsfxsize=.45\textwidth {\epsffile{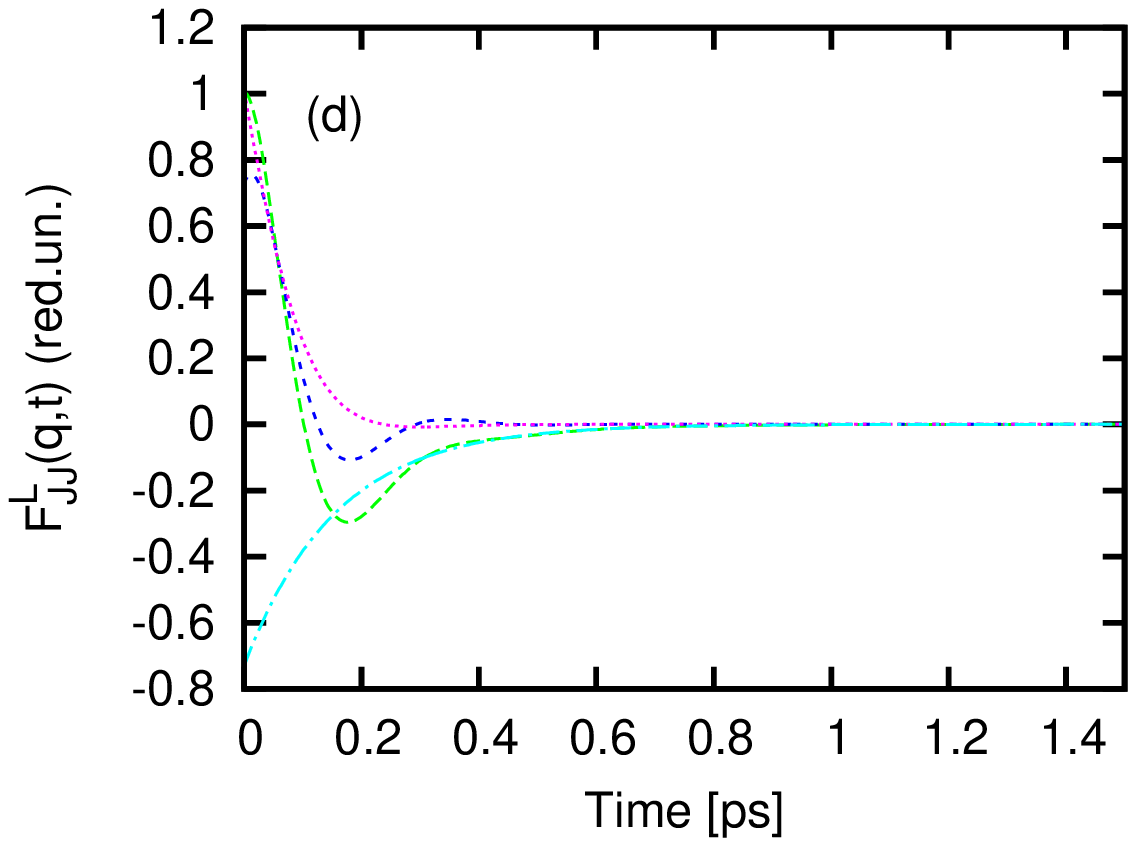}}
\caption{Reproduction of the MD-derived density-density (a) and 
longitudinal current-current (b) time correlation functions by the fit-free 
GCM theory within the five-variable thermo-viscoelastic approach\cite{Bry10} 
for the supercritical Ne at density 1600 kg/m$^3$ and temperature 295~K at 
the wave number $q=2.93 A^{-1}$. Contributions from the same eigenmodes of 
the generalized Langevin equation to the density-density (c) and 
longitudinal mass current-current (d) time correlation functions. 
For the mass current-current functions the reduction by the square of thermal 
velocity $k_BT/m$ and atomic mass $m$ was made.
}
\end{figure}

We performed a fit using the ansatz of \cite{Kry20} but in time domain (see Eq.3 of \cite{Kry19})
$$
F^{tot}_{JJ}(q,t)=mk_BT\large[cos(\omega_L(q))e^{-\Gamma_L(q) t}+2cos(\omega_T(q))e^{-\Gamma_T(q) t}
\large]~,
$$
 in order to avoid 
noise from numerical time-Fourier transformation (see Fig.4) and to compare with the 
fit-free eigenmode contributions shown in Fig.4d. Note, that in \cite{Kry19} the authors
named this fit ansatz as the proper one for analysis of excitation spectra in fluids without 
even a check of that model in comparison with the theory of collective excitations in liquids,
although even in comparison with the damped harmonic oscillator (DHO) their fit ansatz does 
not satisfy the exact short-time sum rule even for the first time derivative.
Obviously, there is no agreement between the 
modes and contributions obtained from fit-free eigenmode theory and the simlified ansatz 
of \cite{Kry20}. Moreover, since exactly the same collective modes contributing to the 
longitudinal mass current-current correlations must contribute to the density-density correlations
$F_{nn}(q,t)$ we tried to fit the two oscillating modes but with amplitudes taken as free 
parameters to the simulation-derived $F_{nn}(q,t)$. Such a fit failed because it was impossible
to recover the MD-derived density-density time correlation function (Fig.3a) with two oscillating 
contributions and neglecting the strong contribution from relaxing eigenmode of 
structural relaxation.
\begin{figure}\label{Fig3S}
\epsfxsize=.45\textwidth {\epsffile{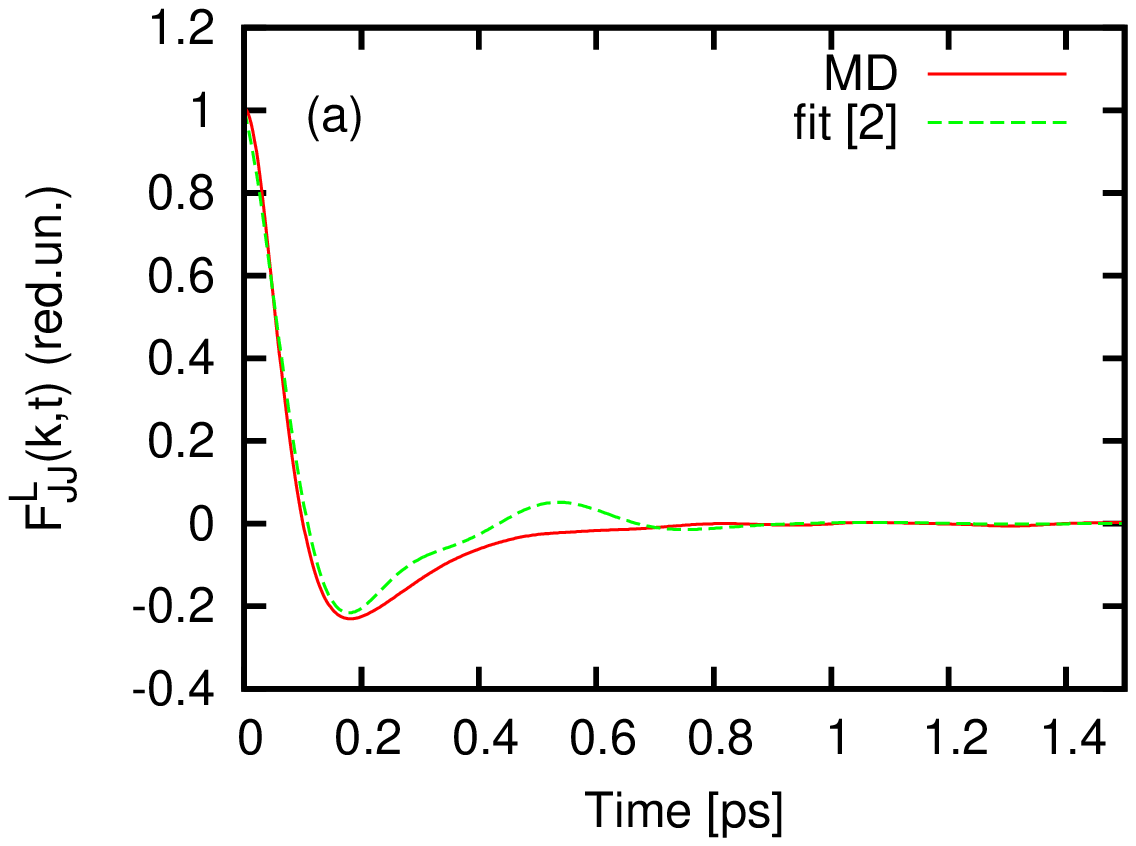}}
\epsfxsize=.45\textwidth {\epsffile{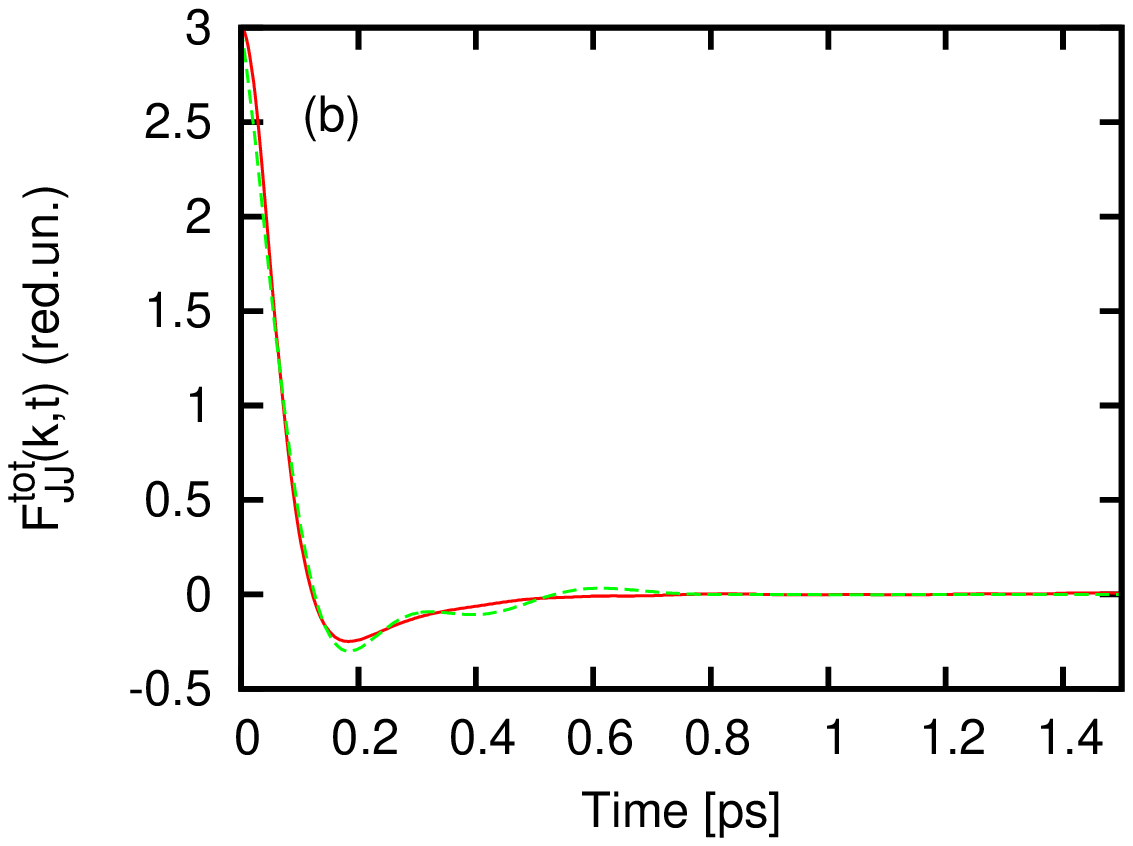}}

\epsfxsize=.45\textwidth {\epsffile{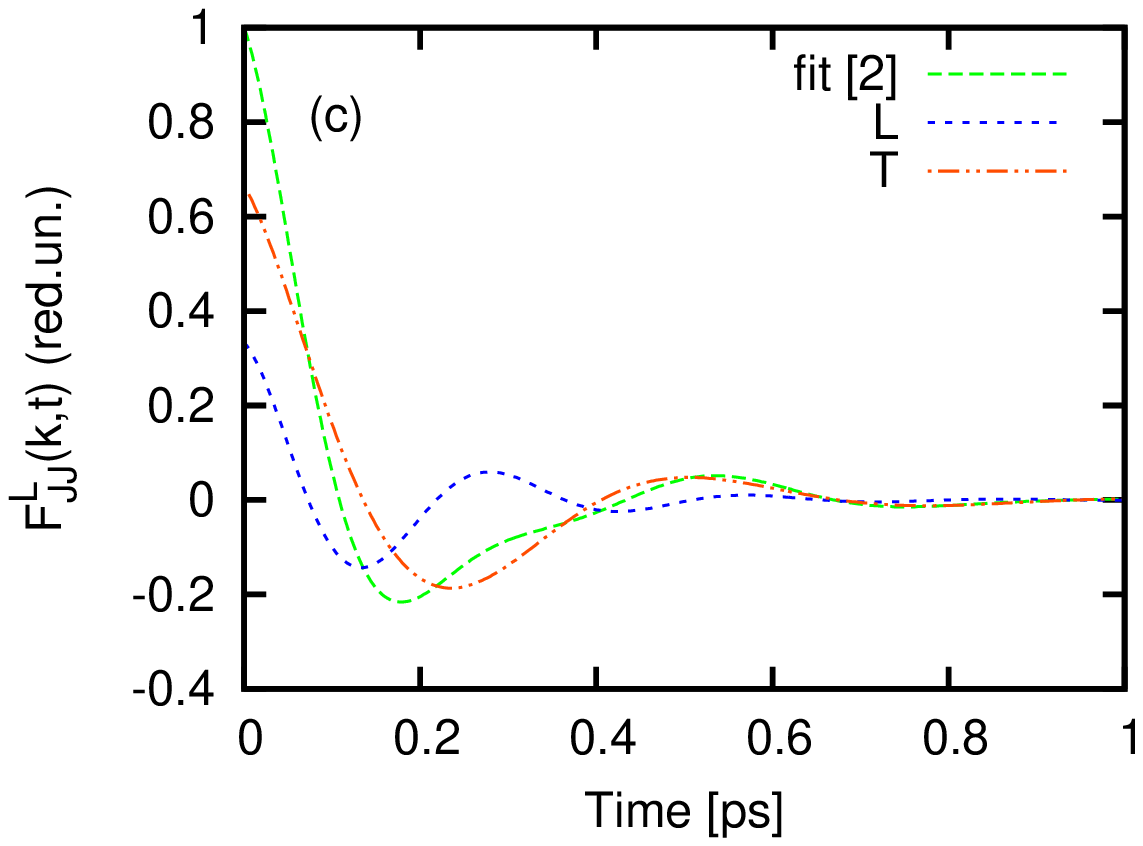}}
\epsfxsize=.45\textwidth {\epsffile{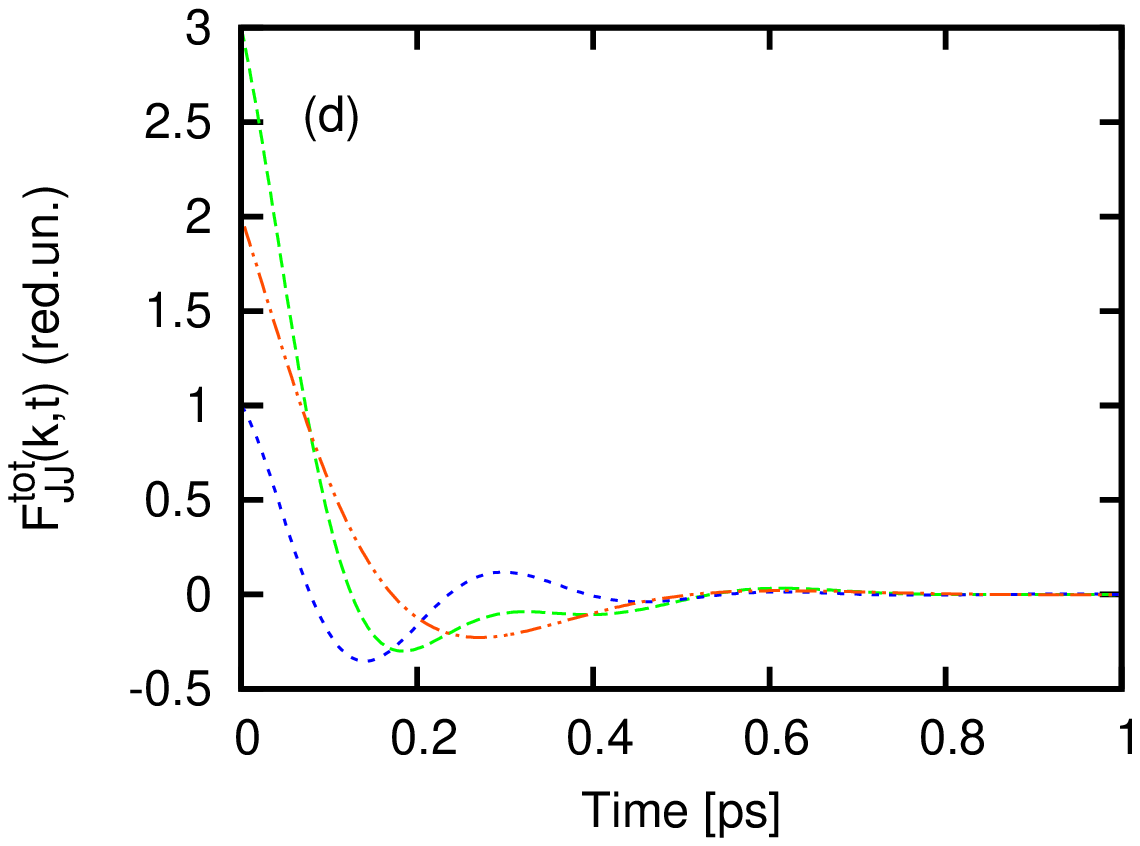}}
\caption{Reproduction of the MD-derived longitudinal (a) and total (b) 
mass current-current time correlation function in time domain by the fit ansatz \cite{Kry20} 
and contributions from L and T excitations to them (c,d) performed for the same thermodynamic
point of supercritical Ne as in Fig.3.
The reduction by the square of thermal velocity $k_BT/m$ and atomic mass $m$ was made.
}
\end{figure}

\end{document}